\documentstyle[epsf]{europhys}

\def\etal{{\hbox{{\tenit\ et al.\/}\tenrm :\ }}}

\def\And{{\rm and\ }}

\def\drm{{\rm d}}

\def\dif#1#2{\frac{\drm#1}{\drm#2}}

\def\stars{\bigskip\centerline{***}\medskip}

\newif\ifboo \boofalse

\renewcommand{\etal}{et al }
\begin{document}
\euro{XX}{X}{X-X}{1998}
\Date{June 12 1998}
\shorttitle{L.\ I.\ PLIMAK  \etal THERMAL PROPERTIES ETC.}
\title{Thermal Properties of Interacting Bose Fields and
Imaginary-Time Stochastic Differential Equations}
\author{L.\ I.\ Plimak\inst{1}, M.\ Fleischhauer\inst{2}
\And D.\ F.\
Walls\inst{1}}
\institute{
     \inst{1} Dept. of Physics, University of Auckland, Private Bag
92019, Auckland, New Zealand\\
     \inst{2} Sektion Physik, Ludwig-Maximilians Universit\"at M\"unchen,
D-80333 M\"unchen, Germany}
\rec{}{}
\maketitle

\begin{abstract}
Matsubara Green's functions for interacting bosons are expressed as
classical statistical averages corresponding to a linear imaginary-%
time stochastic differential equation. This makes direct numerical
simulations applicable to the study of equilibrium quantum properties 
of bosons 
in the non-perturbative regime. 
To verify our results  we discuss an oscillator with quartic anharmonicity as a
prototype model for an interacting Bose gas.
An analytic expression for the characteristic function in a thermal
state is derived and a Higgs-type phase transition discussed, which
occurs when the oscillator frequency becomes negative.
\end{abstract}

\section{Introduction}
A commonly used method
to study thermal properties of interacting many-body
systems is the Matsubara Green's function technique
\cite{Matsb,Abrikosov,Fetter}.
In this paper, we present a method 
of evaluating
Matsubara Green's functions for bosons,
based on constructively 
characterising Feynman paths 
as solutions to an Ito 
stochastic differential equation (SDE). 
(For stochastic calculus 
and SDEs see \cite{Gardiner90}; 
a brief discussion is given below.)
Numerically simulating this SDE 
then allows one to 
calculate physical quantities  
using their expressions as  path integrals 
\cite{Wiegel75} 
(this constitutes a constructive rather than a
Monte-Carlo method \cite{Ceperley95}).
Since path integrals correspond to an in 
principle {\em exact summation\/} 
of the Matsubara 
diagram series as a whole  
(as opposed 
to approximate {\em partial summations\/} 
underlying conventional Green's function 
approaches \cite{Abrikosov,Fetter}),     
the proposed technique is  
applicable beyond the perturbation region 
and is only limited by numerical errors. 
It could be  advantageous, e.g.,  for
analysing the behaviour of an interacting Bose gas near the critical
temperature of condensation \cite{Parkins}.

We develop a version of the Matsubara technique where
a diagram series is averaged over the thermal $P$-distribution
for free bosons \cite{Gardiner91}.
For the grand canonical ensemble  
this distribution is positive and hence
can be interpreted as a probability density.
Then we see that summing the series 
is equivalent to solving a certain SDE.
For bosons with quartic interaction, this SDE
is a linear imaginary-time Schr\"odinger equation with
multiplicative noise.
As a result, we express  
 normally ordered
 thermal averages of the bosonic field 
as classical statistical averages.
We verify this result, including the choice
of stochastic calculus (Ito), by considering a prototype model
of interacting Bose fields---a quantized oscillator
with quartic anharmonicity. A simple analytic expression for the
characteristic function is derived and a Higgs-type phase transition
discussed, which occurs when the frequency of the oscillator
 becomes negative.

\section{General theory}
Consider a system of interacting Bose fields
in
an external potential $V=V(\vec r\,)$,
described by the Schr\"odinger field operator
$\hat \psi=\hat \psi(\vec r\,)$,
with the Hamiltonian $H = H_0 + H _{{\rm\small int}} $, where
\begin{eqnarray}
H_0=\int \drm^3\vec r\, \hat\psi^\dagger
\left(- \frac{1}{2 m}\nabla^2 +
V\right)\hat \psi ,\quad\quad
%
H_{\rm int}=H_{\rm int}(\hat \psi ^{\dagger},\hat  \psi )
=\frac{ \kappa }{2}
\int\! \drm^3\vec r \,
\hat\psi^{\dagger 2}
\hat\psi^2,
\end{eqnarray}
$ \kappa =4\pi a/m$
 and $a$ is the $s$-wave scattering length. We use units where $\hbar=1$.

The aim of the paper is to calculate thermal
averages of normally ordered products of $\hat  \psi $ and
$\hat  \psi ^{\dagger}$ over the grand-canonical
density matrix, 
$\rho ={\rm e}^{(\Omega - H +\mu N)\beta}$.
Here $\beta=1/k_B T$,
$T$ is the temperature, $\mu $  is the chemical potential,
and $Z = {\rm exp}\{-\Omega\beta\} $ is the partition function.
To this end, we
consider the normally-ordered characteristic  functional,
\begin{equation}
\chi(\xi,\xi^*)=\left \langle
{\rm exp}\left ( \xi\hat\psi^\dagger\right )
{\rm exp}\left (-\xi^*\hat\psi\right)\right  \rangle,
\label{gen}
\end{equation}
where 
$\xi\hat\psi^\dagger \equiv \int \drm^3\vec r\, \xi(\vec r\,)
\hat\psi^\dagger(\vec r\,)$ (say) 
and  $\xi = \xi (\vec r\,)$ is an arbitrary
complex function.

To relate averages over $\rho$ to those over the density matrix for the free
system, $\rho_0={\rm e}^{(\Omega_0 - H_0+\mu N)\beta}$, we
introduce the Matsubara interaction picture,
\begin{eqnarray}
\hat\psi_T(\vec r, \tau) &=& {\rm e}^{(H_0-\mu N)\tau}\hat\psi
{\rm e}^{-(H_0-\mu N)\tau},\\
\hat{\tilde\psi}_T(\vec r, \tau) &=& 
{\rm e}^{(H_0-\mu N)\tau}
\hat\psi^\dagger{\rm e}^{-(H_0-\mu N)\tau}= 
\hat{\psi}^{\dag}_T(\vec r, -\tau) ,
\end{eqnarray}
and the thermal ``scattering matrix'', 
\begin{equation}
{\cal S}=T_\tau{\rm exp}\left[ -
{\cal H}(\hat {\tilde \psi}_T, \hat{\psi}_T) \right ]
,\label{S}
\end{equation}
where
${\cal H}(\hat {\tilde \psi}_T, \hat{\psi}_T)
= \int_{-\beta}^0 \drm\tau H_{\rm int}(\hat {\tilde \psi}_T, \hat{\psi}_T)$,
and
$T_\tau$ is the ``time''-ordering operator with respect
to $\tau$.
(Note that we consider Matsubara evolution
from $ \tau = -\beta $ to $ \tau = 0$,
not from $0$ to $\beta $ as usual:
$ \tau = 0$ is the latest, not the earliest, ``time'';
 some of
our relations  therefore
differ slightly from those
in the literature.)
Using $
\rho  =
{{\cal S}\rho_0}/{\langle {\cal S} \rangle_0},
$
where $\langle \dots\rangle_0$ denotes
the ``free'' averaging over
$\rho_0$, we
find with ${\tilde\chi}(\xi,\xi^*)\equiv\chi(\xi,\xi^*)
\, \langle {\cal S}\rangle_0$:
\begin{eqnarray}
{\tilde\chi}(\xi,\xi^*)
=
\left \langle  T_\tau{\rm exp}\left[
-{\cal H}(\hat {\tilde \psi}_T, \hat{\psi}_T)
+ \xi \hat{\tilde \psi}_T(+0) - \xi^*
\hat \psi_T(-0)\right]\right \rangle _0.
\label{gen2}
\end{eqnarray}

Matsubara diagram series \cite{Matsb,Abrikosov,Fetter}
are obtained
by expanding
the exponent in (\ref{gen2})
in a power series,
changing the order of the summation and
averaging, and then factorising the many-operator 
averages into products of free
thermal Green's functions \cite{Fetter,Danielewicz}.
Physical quantities are then obtained  
isolating certain ``leading'' classes of 
diagrams which may be summed yielding Dyson equations. 
This inevitably involves approximations which  
fail if the interaction is not small or in the vicinity 
of a phase transition. 
In order to have an approach which is in principle exact,
we here proceed in a different way.
We first apply Wick's theorem  proper
 so as to bring the 
operator expression in (\ref{gen2}) to normal order,
and then employ the relation,
\begin{eqnarray} 
\label{P0def}
\left \langle :F(\hat \psi^{\dag},\hat \psi):\right \rangle _0
=\int\!\! {\rm D}^2\psi_0\, P\bigl(\psi_0({\vec r})\bigr)
\bigl \langle \psi_0\bigl\vert :F(\hat \psi^{\dag},\hat \psi):\bigr\vert 
\psi_0\bigr\rangle \equiv \overline{\ F(\psi_0^*,\psi_0)\ },
\end{eqnarray}%
Here,  $F$ is an
arbitrary functional of
$\hat\psi(\vec r\,)$ and $\hat\psi^\dagger(\vec r\,)$,
$\int\!\! {\rm D}^2\psi_0$ is a functional integration and 
$P(\psi_0)$ is 
the Glauber $P$-distribution 
for the non-interacting bosons in a thermal state, which is
positive \cite{Gardiner91}.
Wick's theorem can be given a  compact form
using functional derivatives \cite{Akhiezer,Hori}
\begin{equation}
T_\tau F(\hat{\tilde\psi}_T,\hat{\psi}_T)=\
: {\rm e}^{ \Delta }
 F(\tilde \phi,\phi)  \vert_{\phi\to{\hat\psi}_T;
\tilde\phi\to{\hat{\tilde\psi}}_T}:\enspace , \label{wick}
\end{equation}
where
\begin{eqnarray}
\Delta  &=&
\int\!\!\!\int _{-\beta }^0 \drm\tau_1 \drm\tau_2
\int\!\!\!\int \drm^3{\vec r}_1 \drm^3{\vec r}_2
D_T({\vec r}_1,{\vec r}_2;\tau_1- \tau_2)
\frac{\delta^2}{\delta \phi({\vec r}_1,\tau_1)
\delta{\tilde \phi({\vec
r}_2,\tau_2)}},
\end{eqnarray}
$\phi(\vec r\, , \tau )$ and $\tilde \phi(\vec r\, , \tau )$ are arbitrary
independent c-number functions,
and
\begin{eqnarray}
D_T({\vec r}_1,{\vec r}_2;\tau_1 - \tau_2)
= \left \langle 0 \left  |
T_{ \tau }\hat \psi_T({\vec r}_1,\tau_1)
\hat{\tilde \psi}_T({\vec r}_2,\tau_2)
\right  | 0 \right \rangle
=\theta(\tau_1-\tau_2)\left[\hat \psi_T({\vec r}_1,\tau_1),
\hat{\tilde \psi}_T({\vec r}_2,\tau_2)\right]
\label{DT}
\end{eqnarray}
is the Matsubara contraction.
Note that Eq.(\ref{wick}) does not refer to any state vector and that
$D_T$ in (\ref{DT}) is a vacuum and not a thermal
average: 
as opposed to the usual Matsubara 
diagram series the thermal properties
of the non-interacting system are contained in the initial distribution
$P(\psi_0)$ and not in the propagator.
It is straightforward to prove that
at the same time $D_T$
is the retarded Green's function  of the
imaginary-time free Schr\"odinger equation:
\begin{eqnarray}
&&\left(\frac{\partial}{\partial\tau_1}
-\frac{1}{2 m}\nabla_1^2 +V(\vec r_1\,)-\mu\right) D_T(\vec r_1,
\vec r_2;\tau_1-\tau_2)\label{ISE}
=\delta(\tau_1-\tau_2)\delta^3(\vec r_1
-\vec r_2).
\end{eqnarray}
Applying (\ref{wick}) to
(\ref{gen2}),
and using (\ref{P0def}), we find
\begin{eqnarray}
&&{\tilde\chi}(\xi,\xi^*) =
\overline{\,
\Phi \, }, \ \ \Phi =
\label{Psi2}
{\rm e}^{ \Delta }
{\rm exp}\Bigl[-{\cal H}(\tilde \phi,\phi) + \xi \tilde \phi (0) - \xi^*
 \phi(0)\Bigr]\Bigr\vert_{ \phi\to \psi_0;{\tilde \phi}\to \tilde \psi_0}
.
\end{eqnarray}
Here
$\psi_0=\psi_0({\vec r}\, ,\tau)$
and ${\tilde \psi}_0={\tilde \psi}_0({\vec r}\, ,\tau)
={\psi}_0^{*}({\vec r}\, ,-\tau)$ are the
amplitudes of the non-interacting fields
in the coherent state $|\psi_0\rangle$:
$\hat \psi_T({\vec r}\, , \tau )\left |\psi_0 \right \rangle  =
\psi_0({\vec r}\, , \tau ) \left |\psi_0 \right \rangle$
and
$\left\langle \psi_0 \right \vert \hat {\tilde \psi}_T({\vec r}\, , \tau ) =
\left\langle \psi_0 \right \vert\tilde \psi_0({\vec r}\,  ,\tau )$.
For $\tau=0$ the Matsubara interaction picture coincides with the
Schr\"odinger picture hence
$\psi_0({\vec r}\, ,0)=\psi_0(\vec r\,)$ and
$\tilde\psi_0({\vec r}\, ,0)=\psi_0^*(\vec r\,)$.

Our goal is now to express (\ref{Psi2}) as a classical
statistical average.
Since $P(\psi_0) \geq 0$,
this interpretation applies to the averaging over $\psi_0 $
(denoted by the upper bar in (\ref{Psi2})). 
In the quantity $\Phi$
one can  replace
$\xi \tilde \phi (0) $ by $
\xi\psi_0^*$, since
$\tau=0$ is
the largest ``time'' and the exponential
operator  does not act on ${\tilde \phi}(0)$.
Following the Stratonovich-Hubbard
transformation used in path-integral approaches \cite{Negerle},
we introduce a Gaussian stochastic variable
$\eta({\vec r}\,,\tau)$, such that
$\overline{ \eta (\vec r\, , \tau ) \eta ({\vec r\,}^\prime, \tau ') }=
 \kappa\, \delta^3(\vec r\, -{\vec r\,}^\prime) \delta ( \tau - \tau ')$,
and write,
\begin{eqnarray}
{\rm exp}\left[-{\cal H}(\tilde \phi, \phi) \right]
=\overline
{\enspace{\rm exp}\left(
i {\tilde \phi}\eta \phi
\right)\enspace},
\end{eqnarray}
where $
{\tilde \phi}\, \eta\, \phi = \int_{-\beta}^0 \drm\tau\int \drm^3\vec r
\enspace{\tilde \phi}({\vec r\,},\tau)\, \eta({\vec r\,},\tau)\,
\phi({\vec r\,},\tau)$.
We then show that
\begin{eqnarray} %
{\rm e}^{ \Delta }
{\rm exp}  \left [
i \tilde \phi  \eta \phi - \xi ^* \phi(0)
\right ]|
_{\phi \to \psi_0;\tilde \phi \to \tilde \psi_0}
={\rm exp}\left [
i \tilde \psi_0  \eta \psi - \xi ^* \psi(0)
\right ] ,
\label{tobB}
\end{eqnarray}%
where $\psi=\psi({\vec r\,}, \tau )$ is a solution of the Ito equation,
\begin{eqnarray} 
\label{Eqb}
\left(\frac{\partial}{\partial\tau}
-\frac{1}{2 m}\nabla^2 +V-\mu\right)
\psi({\vec r\,}, \tau ) = i  \eta({\vec r\,}, \tau )\psi({\vec r\,}, \tau ),
\label{SDEpsi}
\end{eqnarray}
with the initial condition given by 
\begin{eqnarray}
\psi({\vec r\,},-\beta ) =
\psi_0({\vec r\,},-\beta ) .
\end{eqnarray}
Indeed, expand all exponents
on the LHS of (\ref{tobB}) in power series and perform
the functional differentiations.
The emerging series is similar to a diagram series
for a problem with a linear perturbation.
In particular, all connected diagrams are linear chains,
of the structure either
$\tilde  \psi _0 D_T  i \eta D_T \cdots i \eta  \psi _0$,
or $- \xi ^* D_T  i \eta D_T \cdots i \eta  \psi _0$.
By virtue of Meyer's first theorem \cite{Negerle} this 
immediately results in the RHS of (\ref{tobB}), with $\psi$
 obeying  the Dyson equation,
\begin{equation}
 \psi({\vec r},\tau)  =  \psi _0({\vec r},\tau) + 
\int\!\! \drm^3{\vec r\,}^\prime
\int_{-\beta}^0\!\!\drm \tau^\prime\, 
D_T({\vec r},{\vec r\,}^\prime,\tau-\tau^\prime) 
i \eta({\vec r\, }^\prime,\tau^\prime)  
\psi({\vec r\,}^\prime,\tau^\prime), 
\end{equation}
which in turn is equivalent to (\ref{Eqb}).

For $\eta=0$, Eq.(\ref{SDEpsi}) describes the evolution of
$\psi_0$, while changing $ \tau \to - \tau $ yields the equation 
for $\tilde \psi_0$. Hence
$\frac{\partial}{\partial\tau} \int\! \drm^3{\vec r}\, 
\tilde  \psi _0({\vec r}, \tau ) 
 \psi ({\vec r}, \tau ) =
i\int\! \drm^3{\vec r}\, \tilde  \psi _0({\vec r}, \tau )  
\eta ({\vec r}, \tau ) \psi ({\vec r}, \tau ) $ and we find that   
\begin{eqnarray}    
i\tilde  \psi _0 \eta \psi \equiv
i\int _{- \beta }^0\!\! \drm  \tau \int\!\! \drm^3{\vec r} \,
\tilde  \psi _0({\vec r},  \tau )  \eta ({\vec r}, \tau ) \psi ({\vec r}, \tau )
%
%
= \int\!\! \drm^3{\vec r}\, \psi _0^*({\vec r})\left[ \psi ({\vec r},0) -  
\psi _0({\vec r})\right] . \nonumber  
\end{eqnarray}%
Finally,
\begin{eqnarray} %
 \chi  (\xi ,\xi ^*) =\tilde \chi  (\xi ,\xi ^*) /\tilde \chi  (0,0),
\ \ \tilde \chi  (\xi ,\xi ^*) =
\overline{\overline{\ {\rm e}^{ \Lambda }\ }},
\label{PsiRes}
\\
 \Lambda  =
\int \drm^3 \vec r \left [
\xi (\vec r) \psi _0 ^*(\vec r) - \xi ^* (\vec r)\psi (\vec r, 0)
+ \psi_0^*(\vec r )\psi(\vec r, 0)-|\psi_0(\vec r)|^2
\right ] .
\end{eqnarray}
The double upper bar in (\ref{PsiRes})  
denotes an averaging over
both $ \psi _0(\vec r \, )$ and $ \eta (\vec r \, , \tau )$;
i.e., the statistics here is that of the random
trajectories 
($\equiv$ Feynman paths) 
$ \psi  (\vec r \, , \tau )$, which are solutions to the
SDE (\ref{Eqb}), with the random initial condition
distributed with probability $P(\psi_0)$.
The interaction between the bosons is contained in the noise
$\eta$ while the thermal properties of the system in $P(\psi_0)$. 
Relation (\ref{PsiRes}) is the main result of
this paper.

It is important that
its derivation implies a  {\em causal
regularisation\/} of the function $D_T$.
Namely, $D_T$ (which is a generalised function)
is replaced by a  
certain number of times
continuously  differentiable function, yet preserving
the causality condition $D_T(\vec r, {\vec r\,}^\prime, \tau ) = 0,
\tau < 0$. 
Consider, for example, relation (\ref{Psi2}).  
Expanding all exponents in (\ref{Psi2}) and performing the
functional differentiations results in the quantity $\Phi$
being expressed as a diagram series. 
We  see that the causal regularisation is 
exactly what is necessary to eliminate the ``short-circuited''
diagrams (containing $D_T(\vec r, {\vec r}, 0) $
which is undefined), giving rise to infinities.
These diagrams emerge when the
differential operator $\Delta$ is applied to a single ${\cal H}$,
whereas their absence is required by the normal form
of the interaction (no contractions within the same interaction
Hamiltonian). 
Note that this regularisation makes sense of  
the replacement $\xi \tilde \phi (0) \rightarrow
\xi\psi_0^*$ we used above. 

The causal regularisation of $D_T$  is also 
what leads to the interpretation
of (\ref{Eqb}) as an Ito SDE. 
Since the noise source $\eta(\vec r,\tau)$ in (\ref{SDEpsi})
is singular (delta-correlated white noise), the SDE is mathematically
speaking not defined. 
However the corresponding
integral equation can consistently be interpreted \cite{Gardiner90} 
with a proper
definition of the stochastic increment,  
$
 \psi (\tau)\, \eta(\tau){\rm d}\tau
\equiv 
 \psi (\tau)\,{\rm d}W
$, where 
$
W( \tau ) = \int _{}^{ \tau }  \eta ( \tau ') {\rm d} \tau ' 
$. 
In the Ito stochastic calculus
$ \psi (\tau){\rm d}W$ is understood as  
$ \psi (\tau) [W(\tau +  {\rm d}  \tau )-W(\tau)]$, while
in Stratonovich calculus it is  
$\frac{1}{2}[\psi(\tau )+\psi(\tau +  {\rm d}  \tau )] 
[W(\tau +  {\rm d}  \tau )-W(\tau)]$. 
Alternatively  \cite{Plimak97}, one may replace 
(\ref{Eqb}) by the corresponding integral equation, 
$ \psi  =  \psi _0 + i D_{T} \eta $, which is defined 
given $D_{T}$ is 
regularised. 
It is then easy to see that the causal regularisation of $D_T$ 
results in $ \psi ( \tau )$ being uncorrelated with $ \eta ( \tau )$, 
which is 
the characteristic property of the Ito calculus.
More detailed arguments in favour of this conjecture
are given in \cite{Plimak97}.

It is worth noting that the diagram series for $\Phi$ 
is structurally identical to that for 
Bose-condensed  systems \cite{Abrikosov,Fetter}, 
where the propagator is replaced by  
$D_T$ and the condensate amplitude by the coherent amplitude $\psi_0$. 
Yet, despite this formal similarity, these series are clearly distinct. 
Here the coherent amplitude $ \psi _0 $ applies to all 
bosonic states, not just the condensate. 
This makes it unnecessary to treat condensed 
and non-condensed fractions separately 
as in \cite{Abrikosov,Fetter}. 
A novel feature is furthermore that our series as a whole and not 
the propagator is averaged over the free distribution 
$P(\psi_0)$.
Consequently we find the propagator $D_T$ equal to the vacuum 
average (\ref{DT}), whereas in \cite{Abrikosov,Fetter} 
it is a thermal average.

\section{Quantum oscillator}
To verify relation (\ref{PsiRes}) including 
the said regularisation/calculus conjecture,  
consider an anharmonic oscillator
described by annihilation and creation operators $\hat  \psi $ and
$\hat  \psi ^\dagger$ with
$H_0 =  \omega_0\, \hat  \psi ^\dagger \hat  \psi $ and
$H _{{\rm\small int}} =
 \kappa /2\enspace \hat  \psi ^{\dagger 2}
\hat  \psi ^2$.
The above results apply by
simply dropping the spatial variable,
$ \psi (\vec r, \tau ) \to  \psi ( \tau )$, etc.
Then,
$ \psi _0( \tau ) =  \psi _0 {\rm e}^{- \omega_0  \tau }$,
$\tilde  \psi _0( \tau ) =  \psi _0^* {\rm e}^{\omega_0  \tau }$.
The SDE, which now reads
 \begin{equation}
\left [
 \dif{}{\tau } -  \omega_0 - i \eta ( \tau )
\right ] \psi ( \tau )= 0,\label{SDEosc}
\end{equation}
is readily solved \cite{Gardiner90}, which yields,  
$ \psi (0) = z  \psi_0 $, where $ z   = s {\rm e}^{i\vartheta   } $,
$\vartheta    = \int_{-\beta }^{0}\drm \tau \,  \eta ( \tau )$
is the total random increment, $\overline{\vartheta   ^2} = \beta  \kappa $,
and $s = {\rm e}^{\beta  \kappa /2
}$. This result corresponds to
(\ref{SDEosc}) regarded as an Ito equation. In Stratonovich calculus (say),
one would find $s = 1$.
Using the explicit expression for $P( \psi _0)$
\cite{Gardiner91},  we have ($r = {\rm e}^{\beta  \omega_0 }$),
\begin{eqnarray} %
 \tilde \chi  (\xi ,\xi ^*) = 
\int  \frac{(r-1){\rm e}^{-(r-1)|\psi_0 |^2}\drm^2\psi_0 }{ \pi }
\int  \frac{{\rm e}^{-\vartheta   ^2/2 \beta  \kappa }\drm\vartheta   }
{\sqrt{2 \pi \beta  \kappa }} 
\exp  \left [
-|\psi_0 |^2(1-z) + \xi \psi_0 ^* - \xi ^* z \psi_0
\right ] .
\label{TwoInt}
\end{eqnarray}%
Changing the order of integrations
in (\ref{TwoInt}) and then taking the Gaussian integral over
$ \psi _0$  yields,
\begin{eqnarray} 
 \tilde \chi  (\xi ,\xi ^*) =
\overline{\enspace
 \frac{r-1}{r-z}
{\rm exp}\left (
-  \frac{|\xi |^2 z}{r-z}
\right )
\enspace}=
\frac{r-1}{r}\sum_{n=0}^\infty {\rm L}_n(|\xi|^2)\Bigl(\frac{s}{r}\Bigr)^n
{\rm e}^{-n^2\beta\kappa/2}
,\label{chibar}
\end{eqnarray}%
where the upper bar denotes the integration (averaging)
over $\vartheta $ and ${\rm L}_n$ are the Laguerre polynomials.

To verify the choice of the Ito calculus consider the partition function
of the anharmonic oscillator. Noting that
$\tilde \chi (0,0) = \left \langle {\cal S}\right \rangle _0 =
Z/Z_0$, $Z_0 = 1/(1-  {\rm e}^{-\beta\omega_0}) = r/(r-1)$, 
and using (\ref{chibar}), we have
\label{Z}
\begin{eqnarray} 
Z =   \overline{\enspace \frac{1}{1 - z {\rm e}^{-\beta\omega_0}}
\enspace}=
\sum_{n=0}^{\infty}
 s^n   {\rm exp} \left (
- \beta  \omega_0 n -  \frac{\beta  \kappa n^2}{2}
\right ) .
\label{ZB}
\end{eqnarray}%
This should be compared to the  
 Fock-space expansion, 
$ Z=\sum_{n=0}^\infty {\rm exp}\left\{-\beta  \omega_0 n
-\beta\kappa n(n-1)/2
\right\}.$
We see that the ``Ito-valued''
$s = {\rm e}^{\beta  \kappa /2}$ is exactly what is required
to match (\ref{ZB}) and this expression, while
the ``Stratonovich-valued''
$s = 1$ results in a discrepancy.

In Fig.1, we plot the characteristic function for $\omega_0=1$,
$\beta=0.1$ and different values of the interaction constant.
After an initial decay, $\chi$ exhibits exponentially growing
oscillations.
The $P$-function, which is the Fourier transform of $\chi$,
is hence not
an ordinary function and the thermal state of the interacting
system is {\it non-classical} as opposed to the non-interacting one.

\

\noindent
\begin{minipage}{6.9001cm}
\noindent
\leavevmode
\epsfxsize=8 true cm
\epsffile{matsb_fig1.psf}
\end{minipage}
\\
\begin{minipage}{14.3cm}
\noindent
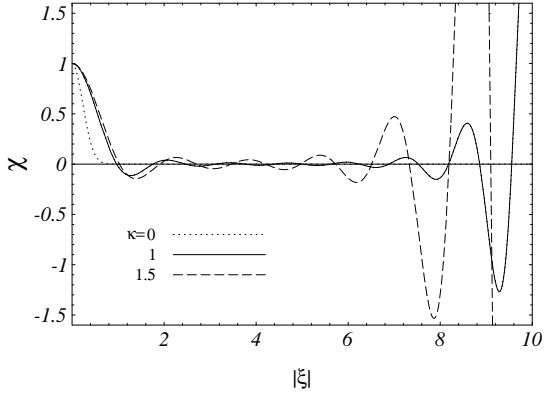
{\small Fig.\ 1. -- Characteristic function of the oscillator for
$\omega_0 =1$, $\beta = 0.1$  and different values of 
anharmonicity.}
\end{minipage}
\\

The anharmonic oscillator shows an interesting behaviour if a
negative linear part is added to the interaction,
$H_{\rm int}=-\Delta\omega\, \hat \psi^\dagger \hat \psi +
\kappa/2\, \hat \psi^{\dagger 2} \hat \psi^2$.
The total
linear part of the Hamiltonian,
$( \omega _0 -  \Delta  \omega )\hat n =  \omega \hat n$, may then be negative.
If $\tilde  \psi ^2 = -  \omega  / \kappa  > 0$,
the  ground state of the interacting system no longer coincides with
that of the harmonic oscillator but with one of the Fock states.
In the limit of infinite system-size, $n_0\equiv
\omega_0 /\kappa\to\infty$, the system undergoes a second-order Higgs-type phase transition,
with $\tilde  \psi $ being the order parameter.
This is illustrated in Fig.2 which shows the normalized 
mean number of quanta $ \nu  =\langle
{\hat n}\rangle/n_0$ and fluctuations $Q=\langle (\hat n - \langle
{\hat n}\rangle)^2\rangle/\langle \hat n\rangle$ as a function of
$\Delta\omega$ for different
system-size parameters.
\\
\noindent
\begin{minipage}{6.9001cm}
\noindent
\leavevmode
\epsfxsize=8 true cm
\epsffile{matsb_fig2.psf}
\end{minipage}
\\
\begin{minipage}{14.3cm}
\noindent
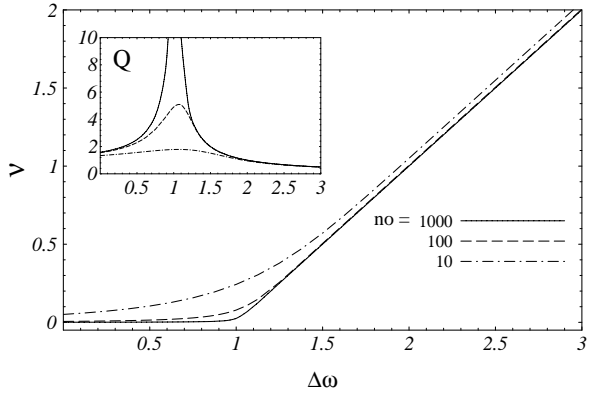
{\small Fig.\ 2. -- Higgs-type phase transition in the anharmonic oscillator.
Mean number of quanta and fluctuations (insert)
for $\omega_0  = 1,\ \beta=1$ and different  $n_0$.}
\end{minipage}
\\

Above threshold,  quantum distribution such as the Wigner function should
become concentrated at $|\psi | \sim
\tilde  \psi $ rather than at $|\psi | \sim 0$.
The above analysis is easily extended to include the additional
linear interaction and all results apply with
$s={\rm e}^{\beta\kappa/2} \to s=
{\rm e}^{\beta\Delta\omega+\beta\kappa/2}$. Thus 
\begin{eqnarray} %
W(\psi ,\psi ^*) = 
\int \frac{\drm^2 \xi}{\pi^2}\, {\rm e}^{\xi \psi ^* - \psi \xi ^* - |\xi |^2/2}
 \chi (\xi ,\xi
^*)
=  \frac{2(r-1)}{ \pi \tilde \chi (0,0)}
\overline{\enspace \frac{1}{r+z}{\rm exp} \left (
- 2 |\psi |^2  \frac{r-z}{r+z}
\right ) \enspace}.
\label{Wigner}
\end{eqnarray}%
The averaging here is over $\vartheta $. 
Rigorously the last equation in (\ref{Wigner}) 
holds only if $\Re(r - z) > 0$ and hence
$ \kappa < 2  \omega $, but one can get 
 rid of this condition by appropriately modifying the averaging  
(namely, moving it from the circle $|z| = s$ to $|z| = 1$ while preserving all $\overline{z^n}, \ n=0,1,\cdots$).
The results of the corresponding numerical evaluation of the Wigner function
shown in Fig.~3 clearly indicate the phase transition when $\omega$ 
becomes negative.
\\
\noindent
\begin{minipage}{6.9001cm}
\noindent
\leavevmode
\epsfxsize=6.9 true cm
\epsffile{matsb_fig3a.psf}
\end{minipage}
\hspace{0.38cm}
\begin{minipage}{6.9cm}
\noindent
\leavevmode
\epsfxsize=6.9 true cm
\epsffile{matsb_fig3b.psf}
\end{minipage}
\\
\begin{minipage}{6.9001cm}
\noindent
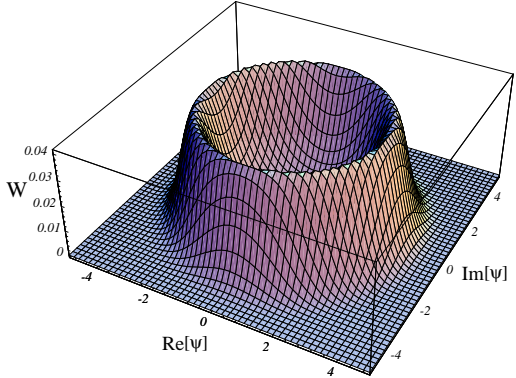
{\small Fig.\ 3a. --  Typical shape of the Wigner function 
``above threshold''
($\beta =0.1, \ \kappa = 1,\ \omega=-8$).}
\end{minipage}
\hspace{0.38cm}
\begin{minipage}{6.9001cm}
\noindent
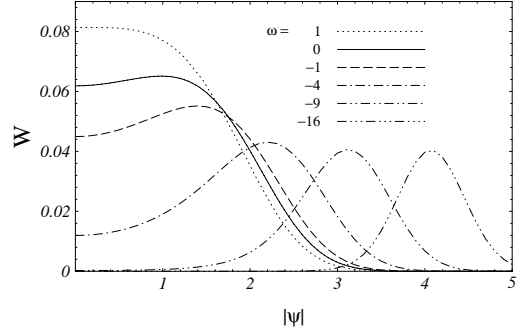
{\small Fig.\ 3b. -- The Wigner function for different values of  $\omega$.}
\end{minipage}
\\ 

In conclusion,  we have shown that in 
the Feynman-path representation of the 
Matsubara Green's function of interacting 
bosons, the Feynman paths may be constructively 
characterised as solutions to an imaginary time 
Ito stochastic differential equation. This 
allows one to
calculate normally ordered thermal averages
of interacting nonrelativistic Bose fields
beyond the level of perturbation.
We have verified this result, 
including the fact that the equation we find is an Ito 
equation,  for the simple prototype model
of an anharmonic oscillator. 
The  relation between stochastic calculus and regularisation
as well as the  application of the method to ultracold atomic quantum 
gases will be subject to further investigations.

%
%
\stars M.F. would like to thank the Physics Department of The
 University of Auckland and Prof. D.F.Walls for  hospitality
during his stay at  Auckland.
This work was supported by the Marsden Fund of the Royal
Society of New Zealand.

%
%

\end{document}